# Consistent thermodynamic derivative estimates for tabular equations of state.

Gary A. Dilts

Continuum Dynamics Group

Los Alamos National Laboratory

Mail Stop D413, Los Alamos NM 87544

ABSTRACT

*Numerical simulations of compressible fluid flows require an equation of state (EOS) to relate the thermodynamic variables of density, internal energy, temperature, and pressure. A valid EOS must satisfy the thermodynamic conditions of consistency (derivation from a free energy) and stability (positive sound speed squared). In many cases an analytic EOS is sufficient, but in many others, particularly when phase transitions are significant, the EOS is complicated and can only be specified in a table. For tabular EOS's such as SESAME from Los Alamos National Laboratory, these can take the form of a differential equation relating the derivatives of pressure and energy as functions of temperature and density, along with positivity constraints. Typical software interfaces to such tables based on polynomial or rational interpolants compute derivatives of pressure and energy and may enforce the stability conditions, but do not enforce the consistency condition and its derivatives. The consistency condition is important for the computation of various dimensionless parameters of an EOS which may involve derivatives up to second order. These parameters are in turn important for the development of more sensitive artificial viscosities and Riemann solvers that accurately model shock structure in regions near phase transitions. We describe a new type of table interface based on the tuned regression method, which is derived from a constrained local least squares regression technique. It is applied to several SESAME EOS's showing how the consistency condition can be satisfied to round-off while computing first and second derivatives with demonstrated second-order convergence. An improvement of 14 orders of magnitude over conventional derivatives is demonstrated, although the new method is apparently two orders of magnitude slower, due to the fact that every evaluation requires solving an 11-dimensional nonlinear system.*

## 1. INTRODUCTION

The two most common techniques for modeling the heat generated by shock waves in the numerical simulation of compressible flows are artificial viscosity [1] and the Riemann solver [2]. Over the years, these have seen many enhancements and variations. For comprehensive summaries, see [3--5]. Unfortunately, the Riemann solver is only well developed for the ideal gas equation of state. Some attempts have been made for more complicated analytic EOS's, such as the Mie-Gruneisen EOS [6--7], but real materials in general have such a complicated EOS that it can only adequately be expressed in a table, for which there is no Riemann solver yet published. The SESAME library [8--9], which is widely used at Los Alamos National Laboratory and has been distributed throughout the world, contains tabular EOS's for many elements and will be the source of our examples. Our treatment in this paper is specific to the choice of variables used in SESAME but it may be straightforwardly modified for others.

The mathematical description of the behavior of shock waves in real fluids with an arbitrary equation of state was described in detail in [10]. Four dimensionless quantities are important:

$$\gamma = -\frac{V}{P}\frac{\partial P}{\partial V}\bigg|_S, \quad \Gamma = -\frac{V}{T}\frac{\partial T}{\partial V}\bigg|_S, \quad g = \frac{PV}{T^2}\frac{\partial T}{\partial S}\bigg|_V, \quad \mathcal{G} = \tfrac{1}{2}\frac{V^2}{\gamma P}\frac{\partial^2 P}{\partial V^2}\bigg|_S. \quad (1)$$

The symbols $P, V, T, S$ represent pressure, specific volume, temperature, and entropy, respectively. The quantity $\gamma$ is the adiabatic exponent, $\Gamma$ is the Grüneisen coefficient, $g$ is the dimensionless specific heat, and $\mathcal{G}$ is the fundamental derivative. The quantities $\gamma$ and $\mathcal{G}$ represent the slope and curvature of isentropes in the $P-V$ plane, respectively. The quantity $\mathcal{G}$ is most important for the determination of shock wave structure. When $\mathcal{G} > 0$, shocks occur in compression; when $\mathcal{G} < 0$, shocks occur in rarefaction. In a numerical simulation this information must be incorporated into the switch used to turn on artificial viscosity or in the solution constructed by a Riemann solver. First and second order approximate Riemann solvers for real EOS's would make extensive use of $\mathcal{G}$. Clearly, in order to construct these solvers we must first know how to compute physically realistic values of $\mathcal{G}$ from tables.

Assume for the moment we have internal energy $E$ expressed as a function of specific volume $V$ and entropy $S$. The thermodynamic definitions of pressure $P$ and temperature $T$ are

$$P = -\frac{\partial E}{\partial V}\bigg|_S, \quad T = \frac{\partial E}{\partial S}\bigg|_V \quad (2)$$

and imply that

$$\frac{\partial P}{\partial S}\bigg|_V = -\frac{\partial T}{\partial V}\bigg|_S. \quad (3)$$

This is the thermodynamic consistency condition and it amounts to a differential equation that a valid equation of state must satisfy. In the SESAME tables, pressure and energy are expressed as functions of temperature and density. With temperature and density independent, condition (3) takes the form

$$P = T \frac{\partial P}{\partial T} + \rho^2 \frac{\partial E}{\partial \rho} \qquad (4)$$

using various thermodynamic identities [10]. Thermodynamic stability requires that the Hessian of $E$ be jointly convex in $V$ and $S$, which leads to the conditions

$$\left.\frac{\partial^2 E}{\partial S^2}\right|_V \geq 0, \quad \left.\frac{\partial^2 E}{\partial V^2}\right|_S \geq 0, \quad \left.\frac{\partial^2 E}{\partial S^2}\right|_V \left.\frac{\partial^2 E}{\partial V^2}\right|_S \geq \left(\frac{\partial^2 E}{\partial S \partial V}\right)^2. \qquad (5)$$

With temperature and density independent these are satisfied if

$$\frac{\partial E}{\partial T} \geq 0, \quad \frac{\partial P}{\partial \rho} \geq 0. \qquad (6)$$

Equations (4) and (6) are thus constraints on any derivatives one might construct from the SESAME table data for $P$ and $E$ in order for them to be physically realistic. They must also be satisfied in the evaluation of the quantities in equation (1). The computation of $\mathcal{G}$ involves second derivatives, so the derivatives of (4) with respect to temperature and density also need to be satisfied to make $\mathcal{G}$ physically realistic. How to do all this is the subject of this paper.

That satisfaction of these constraints is not automatic for traditional derivative evaluation schemes is illustrated by Table 1, which shows the average absolute value of the *log-scale* of the normalized consistency error, $\text{ls}(\varepsilon) \equiv \text{sgn}\,\varepsilon \ln|1+\varepsilon|$, where

$$\varepsilon = \left(-P + T\frac{\partial P}{\partial T} + \rho^2 \frac{\partial E}{\partial \rho}\right) \bigg/ \left(|P| + T\left|\frac{\partial P}{\partial T}\right| + \rho^2 \left|\frac{\partial E}{\partial \rho}\right|\right), \qquad (7)$$

using several common methods [11--12] for computing derivatives of SESAME table 2984 for molybdenum. Since pressure and energy vary by 6 and 12 orders of magnitude respectively in this example the normalizing denominator is necessary for a fair assessment of the error. The table grid was $37 \times 65$, and the evaluation grid was $75 \times 135$. The standard deviation for all methods was approximately 1.0e-15. Thus at most points, traditional derivatives match the consistency condition to a little less than three decimal places. The minimum value of $\partial E/\partial T$ was -0.198 Mbar-cc/K, and of $\partial P/\partial \rho$ was -33.2 Mbar-cc/g using birational derivatives. These are well below the minimum allowed values of zero according to (6). The other methods showed similar results. Some software interfaces have options to enforce these positivity constraints [11], but it is not done in a way which *simultaneously* guarantees satisfaction of the consistency constraints. In the rest of this paper, we elucidate a technique to do precisely this.

**Table 1.**

| Method | Error |
|---|---|
| bilinear | 0.00665 |
| biquadratic | 0.00240 |
| birational | 0.00175 |
| bihermitian | 0.00107 |

Table 1. Average absolute value of the log-scale of the normalized consistency error (7) for various derivative methods in common use. Bilinear, biquadratic, and birational methods are described in [11]. The "bihermitian" method is the bicubic hermitian method described in [12]. The actual software package of [11] was used for the first three methods, and the author's implementation was used for the last.

## 2. NUMERICAL METHODS

The *tuned regression estimator* (TRE) method [13] allows us to estimate derivatives of tabular EOS data while simultaneously guaranteeing (4) and (6). We shall summarize briefly the basic ideas of that paper and refer the reader thereto for more background, generality, detail and examples. Here, we shall just remark that it grew out of the application of the statistical method of local regression estimators [14] to the numerical solution of differential equations.

Let us suppose we have data points in two dimensions $\{y_i\}_{i=1}^N \subset \mathbb{R}^2$ with associated $m$-dimensional data values $\{u_i\}_{i=1}^N \subset \mathbb{R}^m$ which we presume to sample a continuous $m$-valued function of two variables $u : \mathbb{R}^2 \to \mathbb{R}^m$. Suppose we want to estimate $n$ derivatives of $u$ at an arbitrary point $x = (x^1, x^2) \in \mathbb{R}^2$. First, we describe traditional polynomial interpolation methods, in a formalism that will prepare us for local and tuned regression estimation.

Suppose the data points make up a Cartesian grid. Let a column vector of $n$ monomials $p : \mathbb{R}^2 \to \mathbb{R}^n$ be chosen from Table 2. Of course $n$ will be restricted to 4, 9, or 16. The derivative corresponding to a monomial $(x^i)^a (x^j)^b$ is $\partial^{a+b} / (\partial x^i)^a (\partial x^j)^b$. Let $\mathcal{Z}$ be a subset of the data points of size $n$ consisting of the most-centered sub-grid that encloses $x$ of $2 \times 2$, $3 \times 3$, and $4 \times 4$ points for the bilinear, biquadratic, and bicubic methods, respectively. Denote the elements of $\mathcal{Z}$ by $\{z_1, \ldots, z_n\}$ and let $k$ be a map between indices of points in $\mathcal{Z}$ and data points such that $y_{k(i)} = z_i$ for $i = 1, \ldots, n$. We wish to approximate $u(x)$ by $\hat{u}(x) = \zeta p(x)$, where $\zeta$ is a $m \times n$ matrix such that the data are exactly interpolated. Let $Q = [p(z_1), \cdots, p(z_n)]$, an $n \times n$ matrix, and $v = [u_{k(1)}, \ldots, u_{k(n)}]$, an $m \times n$ matrix. We require $\zeta Q = v$, and thus $\zeta = v Q^{-1}$. For any $m$-valued function $f(x)$, let $J_u(x) = [f, \partial f/\partial x_0, \partial f/\partial x_1, \ldots]$ be an $m \times n$ matrix whose columns are the $n$ derivatives corresponding to the monomials in $p$, which we call the *jet matrix*. Then $J_{\hat{u}}(x) = \zeta J_p(x)$. If $u_i = \Lambda\, p(y_i)$, where $\Lambda$ is a constant $m \times n$ matrix, then $v = \Lambda Q$ and $\zeta = \Lambda$, and polynomial interpolants are said to *reproduce* the *basis* $p(x)$. It is well-known that they converge with order $n$ for smooth data, but produce oscillations near discontinuities. The great advantage of polynomial interpolation is speed, as the $Q$ matrix depends only on the $y_i$ and need be computed only once for all $x$.

**Table 2.**

| Method | Monomials |
|---|---|
| bilinear | $[1, x^0, x^1, x^0 x^1]^T$ |
| biquadratic | $[1, x^0, x^1, (x^0)^2, x^0 x^1, (x^1)^2,$ $(x^0)^2 x^1, x^0 (x^1)^2, (x^0)^2 (x^1)^2]$ |
| bicubic | $[1, x^0, x^1, (x^0)^2, x^0 x^1, (x^1)^2, (x^0)^3, (x^0)^2 x^1, x_0 (x^1)^2, (x^1)^3,$ $(x^0)^3 x^1, (x^0)^2 (x^1)^2, x^0 (x^1)^3, (x^0)^3 (x^1)^2, (x^0)^2 (x^1)^3, (x^0)^3 (x^1)^3]$ |

Table 2. Sets of monomials used for several traditional interpolation schemes [11--12]. The biquadratic method here includes 3 extra terms of cubic and quartic order than that described in [11], which might properly be termed the "quadratic" method.

The tuned regression method is a meshfree method and as such, notions of nearness are determined by the value of a real-valued *weight function* $w(x, y_i)$, instead of a grid. The weight function is generally smooth, centrally peaked about $y_i$, and has compact support. When $w(x, y_i)$ is large, $x$ is close to $y_i$. When $w(x, y_i)$ is small, $x$ is far away from $y_i$. Although the grids used in the SESAME tables are non-uniformly spaced Cartesian, meshfree techniques may be applied to them. In this paper, we use the weight function

$$w(x, y_j) = \frac{N_1}{h_j^1} \frac{N_2}{h_j^2} B_4\left(\frac{y_j^1 - x^1}{h_j^1}\right) B_4\left(\frac{y_j^2 - x^2}{h_j^2}\right), \qquad (8)$$

where $B_4$ is the cubic B-spline, defined by

$$B_4(z) = \begin{cases} 1 - \frac{3}{2}z^2 + \frac{3}{4}|z|^3, & |z| \leq 1 \\ \frac{1}{4}(2-|z|)^3, & 1 < |z| \leq 2 \end{cases}$$

and $N_1$ and $N_2$ are constants such that the integral of $W_j$ is 1. We use a vector smoothing length $h_j = [h_j^1, h_j^2]^T$.

Now let $J(x)$ represent the $n \times n$ jet matrix of $p$. It has been verified for a large set of monomial bases that the *shifted basis* has the form

$$p(x, y_i) \equiv J^{-1}(x) p(y_i) = D\, p(y_i - x), \qquad (9)$$

where $D$ is a constant diagonal matrix. Suppose that $\beta(x)$ is a $m \times n$ matrix whose columns are derivative estimates of $u$, the same derivatives that are used in $J(x)$. Then $\beta(x) p(x, y_i)$ is the Taylor series expansion from $x$ to $y_i$. Now suppose that we want $\beta(x)$ to satisfy a set of differential constraints at $x$, say $\mathcal{D}(\beta) = 0$. Through the implicit function theorem, this implies that a subset of the $\beta$ can be eliminated, or equivalently, we can change variables to a smaller number of variables $\gamma$ such that $\beta = \mathcal{E}(\gamma)$ and $\mathcal{D}(\mathcal{E}(\gamma)) = 0$. Our Taylor series then takes the form $\mathcal{E}(\gamma(x)) p(x, y_i)$. The inverse mapping is given by $\gamma = \mathcal{F}(\beta)$ with $\mathcal{E}(\mathcal{F}(\beta)) = \beta$ and $\mathcal{F}(\mathcal{E}(\gamma)) = \gamma$. We measure the average error of the Taylor Series expansion from $x$ to all nearby points $y_i$ with

$$\mathcal{R}(x) = \sum_j \left(u_j - \mathcal{E}(\gamma(x)) p(x, y_i)\right)^2 w(x, y_j). \qquad (10)$$

If we optimize $\mathcal{R}(x)$ with respect to $\gamma(x)$ by solving $\partial \mathcal{R}/\partial \gamma = 0$, we will obtain optimal estimates of $u(x)$ and all of its derivatives through $\beta(x)$. The constraints will be satisfied to round-off by construction: $\mathcal{D}(\beta(x)) \equiv 0$. This constitutes the general method of tuned regression.

The case where $\mathcal{E} = \beta$ is known as the *local regression estimator* (LRE), and has an explicit solution:

$$\beta(x) = \sum_i u_i \psi_i(x)^T, \quad \psi_i(x) = P^{-1}(x) p(x, y_i) w(x, y_i)$$
$$P(x) = \sum_i p(x, y_i) p(x, y_i)^T w(x, y_i) \qquad (11)$$

It is well-studied in the statistics literature [14] and it is easy to show it has the form $\beta(x) = \zeta(x) J(x)$, similar to polynomial interpolants, and it has the reproducing property, just like polynomial interpolants and the moving-least squares (MLS) estimators used in the engineering literature [15]. In fact, the zeroth-derivative estimate of LRE is identical to that of MLS [13]. The convergence rates for $x \in \mathbb{R}$ for the $\nu$-th derivative are $n-v+1$ for $n-v$ odd and $n-v+2$ for $n-v$ even [14]. The moment matrix $P(x)$ in (11) becomes singular when the data points in the neighborhood of $x$ become coplanar, or there are less than $n$ of them, so the smoothing length must be made large enough to prevent these two situations. If it is too large however the procedure becomes expensive, as more neighbors are included in the sums. The proper selection of smoothing length for LRE is a fine art discussed in [14]. It is not known exactly how much of that discussion applies to TRE, but in practice, it is seen that at least $n$ neighbors are also required. It is also wise to monitor condition numbers in the course of solution.

The present application makes use of the following specializations: $x = [T, \rho]^T$, $y_i = [T_i, \rho_i]^T$, $u = [E, P]^T$, and

$$p = \left[1, T, \rho, T^2/2, T\rho, \rho^2/2\right]^T,$$

$$\beta = \begin{bmatrix} E & \partial E/\partial T & \partial E/\partial \rho & \partial^2 E/\partial T^2 & \partial^2 E/\partial T \partial \rho & \partial^2 E/\partial \rho^2 \\ P & \partial P/\partial T & \partial P/\partial \rho & \partial^2 P/\partial T^2 & \partial^2 P/\partial T \partial \rho & \partial^2 P/\partial \rho^2 \end{bmatrix},$$

$$\mathcal{D}(\beta) = \beta_{1,0} - T\beta_{1,1} - \rho^2 \beta_{0,2} = 0, \quad (12)$$

$$\mathcal{F}(\beta) = \left[\beta_{0,0}, \ldots, \beta_{0,5}, \beta_{1,1}, \cdots \beta_{1,5}\right]^T = \gamma,$$

$$\mathcal{E}(\gamma) = \begin{bmatrix} \gamma_0 & \gamma_1 & \cdots & \gamma_5 \\ T\gamma_6 + \rho^2 \gamma_2 & \gamma_6 & \cdots & \gamma_{10} \end{bmatrix} = \beta.$$

We have eliminated $\beta_{1,0}$, which represents the pressure, through $\mathcal{D} = 0$, to define $\gamma$. The evaluation and optimization of (10) is aided by the observation that it can be rewritten as

$$\mathcal{R} = \mathrm{Tr} W + \mathrm{Tr}\left(P \mathcal{E}^T \mathcal{E} - 2 U^T \mathcal{E}\right), \quad (13)$$

where

$$P = \sum_i p_i p_i^T w_i, \quad U = \sum_i u_i p_i^T w_i, \quad W = \sum_i u_i u_i^T w_i \quad (14)$$

and $p_i = p(x, y_i)$, $w_i = w(x, y_i)$.

This prescription addresses the consistency condition (4), but the stability conditions (6) require further attention. We define three possible differential constraints to use:

$$\begin{aligned} \beta_{1,0} - T\beta_{1,1} - \rho^2 \beta_{0,2} &= 0 & \text{(a)} \\ \beta_{0,1} &= 0 & \text{(b)} \\ \beta_{1,2} &= 0 & \text{(c)} \end{aligned} \quad (15)$$

which represent the consistency and stability conditions and adopt a multi-pass algorithm to enforce all constraints simultaneously: (i) First, we try (15)(a) everywhere. (ii)

Where $\beta_{0,1} < 0$ in the result of (i) we apply the combination of (15)(a) and (b). (iii) Where $\beta_{1,2} < 0$ in the result of (i) we apply the combination of (15)(a) and (c). (iv) Where both $\beta_{0,1} < 0$ and $\beta_{1,2} < 0$ in (i), or where $\beta_{1,2} < 0$ in (ii) or $\beta_{0,1} < 0$ in (iii), we apply the combination of (15)(a),(b) and (c). For the SESAME tables examined to date the number of locations where (ii)-(iv) are required is very small. The technique results in the values of $\beta_{0,1}$ and $\beta_{1,2}$ being clamped to zero in regions where passes (i)-(iii) cause them to be negative. In the software, one needs to code four possibilities for $\mathcal{E}$ corresponding to the combinations in passes (i)-(iv). For the four different passes we solve constrained systems with $\{\beta_{1,0}\}, \{\beta_{1,0}, \beta_{0,1}\}, \{\beta_{1,0}, \beta_{1,2}\}, \{\beta_{1,0}, \beta_{0,1}, \beta_{1,2}\}$ eliminated respectively, which results in solving $11 \times 11, 10 \times 10, 9 \times 9$ systems, respectively. This illustrates a key feature of tuned regression: by eliminating some derivatives, we solve a smaller system with enhanced accuracy. The penalty is that the smaller system is more complicated.

Both local regression and tuned regression in regions where only the consistency constraint is enforced produce approximations with smoothness equal to that of the weight function.

A Mathematica [16] program has been written to symbolically optimize (13) for arbitrary $P$, $U$, $W$ and $\mathcal{E}$. It then generates code which is spliced into a C++ library called LORELI (LOcal REgression LIbrary) which is used to operate on the SESAME data. The appendix exhibits the expressions for $\mathcal{R}$ generated by this Mathematica program in terms of the matrices in (14) when the various combinations of constraints in (15) are active.

### 3. EXAMPLE: ANALYTIC

Let $J_u = [u, \partial_T u, \partial_\rho u, \partial_T^2 u, \partial_T \partial_\rho u, \partial_\rho^2 u]$, and suppose that $J_u = \Lambda J$, and $\mathcal{D}(J_u) = 0$. That is, $u$ is a quadratic function that exactly satisfies the consistency and stability conditions. Such an example is given by

$$u = \begin{bmatrix} E \\ P \end{bmatrix} = \begin{bmatrix} -1 + T + \rho + T^2 \\ -T + T\rho + \rho^2 \end{bmatrix} \quad (16)$$

with

$$\Lambda = \begin{bmatrix} -1 & 1 & 1 & 2 & 0 & 0 \\ 0 & -1 & 0 & 0 & 1 & 2 \end{bmatrix}. \quad (17)$$

Now suppose that $u_i = u(y_i) = \Lambda p(y_i)$. Then

$$\mathcal{R}(x) = \sum_i \left( \Lambda p(y_i) - \mathcal{E}(\gamma(x)) J^{-1}(x) p(y_i) \right)^2 w(x, y_i) \quad (18)$$

is minimized if $\gamma = \mathcal{F}(J_u)$, because $\mathcal{E}(\gamma(x)) = J_u(x) = \Lambda J(x)$ and $\mathcal{R}$ is identically zero. In other words, tuned regression possesses the reproducing property just like polynomial interpolation and local regression: a polynomial solution of the differential constraints evaluated at discrete points $\{y_i\}$ will be exactly reproduced at an arbitrary $\{T, \rho\}$, to round-off. and can be used by a software implementation as a verification tool. The LORELI library mentioned above has been so checked on this example and does indeed reproduce to round-off. On the basis of these results we surmise that the tuned regression method for SESAME data is at least third-order accurate. There is a formal proof, but its presentation is out of scope here, however numerical examples below will confirm it.

## 4. EXAMPLE: OXYGEN

We choose SESAME table 5011 for Oxygen at low temperatures as our first example because the 23x51 grid is fairly uniform. Most SESAME tables have grids that are exponential in character to handle the many orders of magnitude variation of temperature and density required. This leads to ill-conditioned matrices in the TRE solution process, and requires special treatment as described below. Table 5011 however does not present this problem. On a 45x103 grid TRE agrees well with the input data and produces no discernable difference to the eye. Figure 1 shows the pressure derivative with respect to density, and a prominent feature is the flat annulus at low temperature. This is a region where the stability constraints were active, and the algorithm did what it was supposed to. The zeroth derivative estimates were not seen to echo this feature to the eye, which illustrates how the derivatives are estimated independently in TRE. The estimate of the derivative is not the derivative of the estimate, as it is in finite element or spectral methods. The two do converge however as the data become dense and the smoothing length goes to zero. The value of the un-normalized consistency error was everywhere less than 1.e-13, which is close to round-off, as promised. Figure 2 shows the relative error between the TRE result and the input table values when the input and output grid are identical. There is good agreement except at low temperatures where the constraints become active.

## 5. LOGARITHMIC FORM

As mentioned above, the exponential grids present in many SESAME tables present numerical difficulties, so a method must be devised to treat the ill-conditioned matrices $\partial^2 \mathcal{R}/\partial \gamma^2$ that appear in the Newton solver for the equations $\partial \mathcal{R}/\partial \gamma = 0$. These occur because the wide range of powers that appear in the moment matrix $P(x)$ in (13) get drastically out of balance when applied to very large numbers. For example, assume 8 decades of range in table coordinates and 45 points, which gives a ratio of about 1.5 in the size of successive intervals. Adjusting $h$ so that there are 49 neighbor points it is easy to verify the condition number of $P \approx 10^{18}$ when $x \approx 10^4$ and $\approx 10^{25}$ when $x \approx 10^6$. One way to restore good conditioning is to use a preconditioner in the solver, and this is under investigation. Another way is to logarithmically scale the variables. To scale the independent variables we use the following transformation:

$$\tau = \ln T, \quad r = \ln \rho, \quad \varepsilon = e\rho. \tag{19}$$

In terms of these variables, the consistency and stability conditions become

$$P + \varepsilon = \frac{\partial P}{\partial \tau} + \frac{\partial \varepsilon}{\partial r} \quad (a)$$

$$\frac{\partial \varepsilon}{\partial \tau} \geq 0 \quad (b) \tag{20}$$

$$\frac{\partial P}{\partial r} \geq 0 \quad (c)$$

These we refer to as *semi-log* constraints. It is the simplest form of the consistency condition and is linear, just like the original consistency condition, and thus requires only one 11x11 solve at most points. This coordinate change was found to work for a few tables, but a further step was required to get satisfactory behavior, because even though the new grid is not exponential, the data for pressure and energy has become exponential, and the problem of ill-conditioned matrices still appears. We now transform the dependent variables by means of

$$\varepsilon_s = \min \varepsilon - \varepsilon_0, \quad p_s = \min P - P_0,$$
$$\eta = \ln(\varepsilon - \varepsilon_s), \quad \zeta = \ln(P - P_s). \tag{21}$$

The new energy and pressure minima $\varepsilon_0$ and $P_0$ are arbitrary but must be positive. In the rest of this paper we set them equal to 1 so that the quantities $\zeta$ and $\eta$ have a minimum value of zero and are always positive otherwise. The larger we make $\varepsilon_0$ and $P_0$ the flatter the transformed data surfaces become in the large. The consistency and stability conditions for $\zeta$ and $\eta$ are

$$\exp(\zeta)\left(\frac{\partial \zeta}{\partial \tau} - 1\right) + \exp(\eta)\left(\frac{\partial \eta}{\partial r} - 1\right) = p_s + \varepsilon_s \quad (a)$$

$$\frac{\partial \zeta}{\partial r} \geq 0 \quad (b) \tag{22}$$

$$\frac{\partial \eta}{\partial \tau} \geq 0 \quad (c)$$

Notice now that the consistency constraint is nonlinear, whereas previously it was linear. We refer to these as *log-log* constraints. We use the LRE solution (11) as the initial condition for a Newton solver or steepest-descent solver to optimize $\mathcal{R}$. In practice, we typically see convergence in 3-5 Newton iterations with this initial condition. The same four-pass strategy for enforcing the stability constraints applies as was used before.

The LORELI library thus contains twelve separate encodings of the residual function and its derivatives: for each type of constraint (flat, semi-log, and log-log), there are four versions corresponding to the combination of consistency and stability constraints listed in the four passes following equation (15). The appendix exhibits these residual functions.

The expressions for the dimensionless derivatives when logarithmic transformations are employed are given by:

$$\gamma = \frac{e^{\zeta}\left(e^{\zeta-\eta}\zeta_\tau^2 + \zeta_r \eta_\tau\right)}{\left(p_s + e^{\zeta}\right)\eta_\tau},$$

$$\Gamma = \frac{e^{\zeta-\eta}\zeta_\tau}{\eta_\tau},$$

$$g = \frac{e^{-\eta}\left(p_s + e^{\zeta}\right)}{\eta_\tau},$$

$$\mathcal{G} = e^{-\eta}\left(e^{2(r+\zeta)}\left(2\eta_\tau\zeta_\tau - \eta_{\tau\tau}\right)\zeta_\tau^3 + 3e^{r+\zeta+\eta}\zeta_\tau\zeta_{\tau r}\eta_\tau^2 + \right. \tag{23}$$

$$e^{2\eta}\left(\zeta_r + \zeta_r^2 + \zeta_{rr}\right)\eta_\tau^3 - e^{\zeta}\zeta_\tau^2\eta_\tau\left(-2e^{2r+\zeta}\zeta_{\tau\tau} + \right.$$

$$\left(e^{2r+\zeta} - 2e^{\eta} + 2e^{r+\eta} + 2e^{r}e_s + e^{2r}p_s - 3e^{r+\eta}\zeta_r\right)\eta_\tau +$$

$$\left.\left.e^{r+\eta}\eta_{\tau r}\right)\right)\Big/\left(2\eta_\tau^2\left(e^{\zeta}\zeta_\tau^2 + e^{\eta}\zeta_r\eta_\tau\right)\right)$$

The reciprocal factors of $\eta_\tau$ and $\eta_\tau^2$ become significant if the specific heat becomes small as discussed below. The expression for $\mathcal{G}$ makes use of the consistency constraint (22a). One could further incorporate its derivatives with respect to $\tau$ and $r$:

$$e^{-\tau}\left(e^{\zeta}\left(\zeta_\tau - \zeta_\tau^2 - \zeta_{\tau\tau}\right) - e^{\eta}\left(\left(\eta_r - 1\right)\eta_\tau + \eta_{\tau r}\right)\right) = 0$$

$$e^{-r}\left(-e^{\zeta}\left(\zeta_r\left(\zeta_\tau - 1\right) + \zeta_{\tau r}\right) - e^{\eta}\left(-\eta_r + \eta_r^2 + \eta_{rr}\right)\right) = 0$$

but the derivation of the log-log TRE method would have to be modified to add these two equations as constraints to those of (22). As it stands, the expression for $\mathcal{G}$ in (23) is consistent with the log-log TRE method implied by (22) above.

### 6. EXAMPLE: MOLYBDENUM

The log-log TRE method was applied to SESAME table 2984 for Molybdenum and the results are shown in Fig. 3. The input grid size was $37 \times 65$, the output grid was $75 \times 135$, and once again there was no discernable difference to the eye between the two. The normalized log-log consistency error, given by

$$\left(e^{\zeta} + e^{\eta} + \varepsilon_s + p_s - e^{\zeta}\zeta_\tau - e^{\eta}\eta_r\right)\Big/ \tag{24}$$

$$e^{\zeta} + e^{\eta} + |\varepsilon_s| + |p_s| + e^{\zeta}|\zeta_\tau| + e^{\eta}|\eta_r|$$

was smaller in magnitude than 2e-16 at all points as required and is 14 orders of magnitude smaller than the values of Table 1 obtained by traditional derivatives, a substantial improvement. The condition number of the final iteration of the Newton solver at each evaluation point was everywhere less than $10^5$, and the number of Newton iterations required to converge to a tolerance of 1.e-13 was at all points between 1 and 4 iterations, a very reasonable number for such a nonlinear problem. The condition numbers are rather high, but it appears possible to reduce them considerably by a simple scaling procedure which may be reported in subsequent publications.

The dimensionless quantities of (23) were computed using the TRE derivatives. At high temperatures, the values of $\gamma$ and $\mathcal{G}$ approach the theoretical values for monatomic ideal gases of 5/3 and 4/3, respectively, which gives some confidence to the calculations. On the other hand, at lower temperatures, there seem to be rather large divergent regions which correlate well with the flat regions at lower temperature in the plot of $\partial \eta / \partial \tau$ in Fig. 4 which have low values ($\leq 1.e-03$). This is significant because the expressions for $\gamma$, $\Gamma$ and $g$ in equations (23) have $\partial \eta / \partial \tau$ in the denominator, and $\mathcal{G}$ has $(\partial \eta / \partial \tau)^2$. To test that this is the cause of the divergent regions, the dimensionless derivatives are multiplied by the appropriate power of $\partial \eta / \partial \tau$ and plotted in Fig. 5. The divergent behavior has been mostly eliminated and in the plot of $\gamma$ one can see the outline several phase boundaries. At low $T$ and high $\rho$ we expect to find the solid phase, and at low $T$ and low $\rho$ we expect to find the mixed phase. In these two phases, the theory leading to the definition of the dimensionless derivatives is incomplete because it does not include the effects of deviatoric strains or stresses and thus nonsensical results may be inescapable. Also, there are known jump conditions on $\gamma$ and $\Gamma$ that may come into play across phase transitions, and these have not been enforced. One of the main points of all this calculation is the determination of the sign of $\mathcal{G}$, which is negative mostly in the mixed phase region, so the ability to provide reliable guidance to numerical methods for shock waves in these regions is clouded. In light of these observations, it seems an appropriate approach is to include the explicit phase boundaries in the EOS evaluator and make reasonably correct estimates of the dimensionless derivatives when they are crossed. It is also possible that some of the divergent behavior seen with TRE is caused by numerical difficulties in early iterations of the Newton solver, and this should be investigated.

Figure 6 shows the $\gamma$ calculated by birational method of the EOSPAC library [11], which does not contain any divergent regions like the TRE result. Figure 7 shows a comparison of $\gamma$ by birational and TRE at three different temperatures, roughly 2 eV, 1 keV, and 100 keV. In general, for positive $\ln \rho$, the two agree fairly well, except at the high-density boundary of the table. At the highest temperatures, both results approach the theoretical value for an ideal gas. For negative $\ln \rho$, the two results are always in disagreement. Perhaps this is because the constraints are more active in that region. This too should be further investigated.

To compare the computational cost of each method, a grid of 750x1350=1012500 points was constructed. The EOSPAC birational evaluation on a 1.7GHz Pentium IV system took 5.7 sec. The time for LRE was 172 sec, which involved neighbor finding using a general two-dimensional binning algorithm and solving and performing one 6x6 linear solve. The log-TRE method doing a single 11x11 linear solve took 249 sec, and the log-log-TRE method doing multiple 11x11 solves took 634 sec, which are 40 and 100 times slower than EOSPAC, respectively. This is disappointing, but not unexpected because in addition to the linear solves involved, the algebra for TRE is much more complicated than for EOSPAC or even LRE. In practice, this computational cost can be avoided by evaluating all derivatives of an EOS on a fine grid and storing them for later evaluation by normal means of interpolation, such as LRE. Presumably the consistency condition would not be violated too much. This too needs further investigation. It must be observed that a linear TRE formulation is possible which would guarantee the consistency condition and involve only a 5x5 solve. Presumably, this would be competitive with LRE and EOSPAC from a performance viewpoint, but one would not be able to use it to compute the fundamental derivative, which requires quadratic TRE at a minimum.

The LORELI implementation of log-log TRE was applied to tables for copper (3333), aluminum (3719), and tin (2160) with similar results, except for some anomalous divergences in one corner which seem to be due to poor choice of smoothing length. When the smoothing length is too small, there are too few neighbors, and the LRE or TRE methods develop ill-conditioned matrices. It becomes an issue near a table boundary because there are fewer neighbors than in the interior. If the smoothing length is too large, it may not be possible to satisfy the constraints with finite values. For LRE there is a fairly well-developed methodology for choosing the smoothing length, but more research is needed to do the same for TRE.

The molybdenum table 2984 does not appear to contain Maxwell constructions for the removal of van der Waals loops. When the log-log TRE method is applied to tables that seem to have Maxwell constructions, such as gold (SESAME 2700), there are severe convergence problems in the vicinity thereof. This may be because second-order interpolation methods generally sustain oscillations in the vicinity of discontinuities, and perhaps because of the exponentials in the TRE method, these oscillations cause serious ill-conditioning in the solvers, both Newton and steepest descent. Some tables seem to have apparently arbitrary abrupt transitions at the edges which also cause a similar problem. These tables may require more physical adjustments at the edges before the TRE method is robust on them. Clearly the issue of Maxwell constructions requires more research. Perhaps they can be detected, by a linear LRE estimate for example, which always seems to be monotone (although a proof is unknown to the author), and then a separate technique applied. This is left for future investigations.

# 7. CONVERGENCE AND ACCURACY

To test convergence of the TRE method, we use the following biquartic EOS

$$E = \left(\rho^3 + \rho^2 + \rho + 1\right)\left(T^4 + T^3 + T^2 + 1\right) + T$$
$$P = -\tfrac{1}{6}\rho^2\left(3\rho^2 + 2\rho + 1\right)\left(2T^4 + 3T^3 + 6T^2 - 6\right)$$
(25)

which satisfies the consistency and stability constraints. It is not reproducible by either quadratic or cubic polynomial, LRE, or TRE methods. This EOS was sampled on a grid of $17 \times 17$ points centered at $(0.5, 0.5)$ with spacing $\Delta y = 2^{-4-k/2}$, $k = 0,\ldots,18$ in each dimension to generate a table of energy and pressure that was input to the various estimation procedures operating on $5 \times 5$ grid centered at $(0.5, 0.5)$ with spacing $\Delta y/4$. The error in the estimates was measured and two sample results for pressure are plotted in Fig. 8. The curves have been truncated on the left where convergence ceased for each method. In particular, the linear methods were the most robust (working at smaller mesh spacings), followed by the quadratic and cubic methods. The bicubic hermitian method was the least robust. Table 3 shows the convergence rates for the various methods coded by the author. The local regression methods converge in keeping with the theoretical rates given in section 2 above. The tuned regression zeroth derivatives converge at the same rates as the quadratic local regression estimator, which is encouraging. For zeroth derivatives the convergence is fourth order, which is remarkable since only a quadratic polynomial is used in the modeling. Table 4 shows the $\log \Delta y = 0$ intercept of the convergence curves, which gives an indication of the relative accuracy of the various methods. The various regression methods trade advantages in different derivatives with their competitors of like polynomial order. The bihermitian method, as coded by the author, seems to have a markedly higher intercept than the other cubic methods, implying that a finer table is required to get the same accuracy as a bicubic or cubic LRE method could get. The tuned regression estimator does the best job on consistency of course, as evidenced by the intercept value. The other methods all converge in consistency error, as they must if they converge at all, but unless you build consistency into the algorithm, you can not guarantee it.

**Table 3.**

|  | lre1 | lre2 | lre3 | tre | bilinear | biquad | bicubic | biherm |
|---|---|---|---|---|---|---|---|---|
| $E$ | 2 | 4 | 4 | 3.9 | 2 | 3 | 4 | 4.3 |
| $P$ | 2 | 4 | 4 | 4.1 | 2 | 3 | 4 | 4.4 |
| $\partial E/\partial T$ | 2 | 2 | 3.9 | 2 | 1 | 2 | 3 | 3.3 |
| $\partial P/\partial T$ | 2 | 2 | 4 | 2 | 1 | 2 | 3 | 3.4 |
| $\partial E/\partial \rho$ | 2 | 2 | 3.9 | 2 | 1 | 2 | 3.9 | 3.4 |
| $\partial P/\partial \rho$ | 2 | 2 | 4 | 2 | 1 | 2 | 3 | 3.6 |
| $\partial^2 E/\partial T \partial \rho$ |  | 2 | 2 | 2 | 1 | 2 | 2.9 | 2.2 |
| $\partial^2 P/\partial T \partial \rho$ |  | 2 | 2 | 2 | 1 | 2 | 2.9 | 2.4 |
| Consistency | 2 | 2 | 4 |  | 1 | 2.1 | 3.1 | 3.4 |

Table 3. Convergence rate for various method on biquartic EOS (25). Linear local regression can not compute second derivatives accounting for vacancies in column 2. Tuned regression gets consistency to round-off, so no convergence figure appears in its column. "Biherm" denotes bicubic hermitian.

**Table 4.**

|  | lre1 | lre2 | lre3 | tre | bilinear | biquad | bicubic | biherm |
|---|---|---|---|---|---|---|---|---|
| $E$ | 0.4 | 0.4 | 0.9 | 0.4 | 0.4 | 0.4 | 0 | 1.2 |
| $P$ | 0.2 | 0.6 | 1 | 0.4 | 0.3 | 0.4 | 0.1 | 1.4 |
| $\partial E/\partial T$ | 0.9 | 1 | 1 | 1 | 0.9 | 0.8 | 0.6 | 2.8 |
| $\partial P/\partial T$ | 0.6 | 0.8 | 1.3 | 0.6 | 0.3 | 0.1 | -0.2 | 3.3 |
| $\partial E/\partial \rho$ | 0.6 | 0.8 | 0.8 | 0.9 | 0.6 | 0.3 | 0.1 | 3 |
| $\partial P/\partial \rho$ | 0.8 | 0.7 | 1.5 | 0.7 | 0.9 | 0.9 | 0.7 | 3.2 |
| $\partial^2 E/\partial T \partial \rho$ | 0.9 | 1.1 | 1.4 | 1.1 | 1.3 | 1.1 | 0.7 | 2.3 |
| $\partial^2 P/\partial T \partial \rho$ | 0.7 | 1.3 | 1.6 | 1.3 | 1.4 | 1.2 | 1 | 2.9 |
| Consistency | 0.5 | 0.2 | 0.1 | -15.5 | 0.1 | -0.1 | -0.3 | 3 |

Table 4. Extrapolated $\log \Delta y = 0$ intercept of convergence curves similar to those of Fig. 8. For methods of like order of convergence, these figures indicate relative accuracy. "Biherm" denotes bicubic hermition.

## 8. MESHFREE ILLUSTRATION

Finally, in Fig. 9(a), we show the molybdenum table 2984 sampled at 21583 random points uniformly distributed across the full range of temperature and density using the log-log TRE of section 5 with $\varepsilon_S = P_S = -10^3$, which flattens the graphs of $\eta$ and $\zeta$ compared to Fig. 3. Each point is colored with the value of $\zeta$ obtained from TRE. This data is then used as an input "table" for resampling at a finer uniform random distribution of 64749 points, again using log-log TRE. The results are plotted in Fig. 9(b) and exhibit the expected behavior. The consistency error was zero to fifteen decimal places as with the other TRE examples. The significance of the meshfree nature of TRE for EOS data is that traditional rectangular-grid tables may be supplemented with extra data points near Maxwell construction or phase change boundaries to achieve greater resolution near these discontinuities with no loss of consistency or accuracy.

## 9. CONCLUSION

We have shown that traditional numerical derivatives of equation of state tables do not simultaneously satisfy the thermodynamic consistency and stability conditions, and that a tractable method to do so can be developed from the tuned regression estimator. The LORELI implementation has demonstrated the reproducing property for quadratic analytic EOS's which satisfy the consistency and stability conditions. Trials on a few SESAME tables have shown that the consistency and stability constraints can be simultaneously enforced to round-off without sacrificing accuracy and that theoretical values for $\gamma$ and $\mathcal{G}$ are approached at high temperature. Versions of the theory and software were developed for flat, semi-log, and log-log coordinates, the last being necessary to handle tables with exponential grids. The convergence rates follow those of the statistical local regression estimator, giving fourth order for zeroth derivatives, and second order for first and second derivatives. The TRE method is apparently much more expensive than traditional derivatives, however. The meshfree character of the TRE method was convincingly demonstrated and provides a basis for augmenting conventional tables with extra data points near physical discontinuities. There are outstanding issues regarding phase boundaries, Maxwell constructions, and table-edge drop-offs which require further research before the technique can be made into a fully robust tool.

## APPENDIX: RESIDUAL EXPRESSIONS

The residual for the consistency constraint given in (15a) is

$$\frac{W_{11}}{2} + \frac{W_{22}}{2} + \gamma_0 \left(-U_{11} + \frac{P_{11}\gamma_0}{2} + P_{12}\gamma_1 + P_{13}\gamma_2 + P_{14}\gamma_3 + P_{15}\gamma_4 + P_{16}\gamma_5 \right) +$$

$$\gamma_1 \left(-U_{12} + \frac{P_{22}\gamma_1}{2} + P_{23}\gamma_2 + P_{24}\gamma_3 + P_{25}\gamma_4 + P_{26}\gamma_5 \right) +$$

$$\gamma_3 \left(-U_{14} + \frac{P_{44}\gamma_3}{2} + P_{45}\gamma_4 + P_{46}\gamma_5 \right) + \gamma_4 \left(-U_{15} + \frac{P_{55}\gamma_4}{2} + P_{56}\gamma_5 \right) + \gamma_5 \left(\frac{P_{66}\gamma_5}{2} - U_{16} \right) +$$

$$\gamma_2 \left(-U_{21}\rho^2 + P_{13}\gamma_7\rho^2 + P_{14}\gamma_8\rho^2 + P_{15}\gamma_9\rho^2 + P_{16}\gamma_{10}\rho^2 - U_{13} + \left(\frac{P_{11}\rho^4}{2} + \frac{P_{33}}{2}\right)\gamma_2 + P_{34}\gamma_3 +\right.$$

$$\left. P_{35}\gamma_4 + P_{36}\gamma_5 + (T P_{11}\rho^2 + P_{12}\rho^2)\gamma_6 \right) + \gamma_6 \left(-T U_{21} - U_{22} + \left(\frac{P_{11}T^2}{2} + P_{12}T + \frac{P_{22}}{2}\right)\gamma_6 +\right.$$

$$\left. (T P_{13} + P_{23})\gamma_7 + (T P_{14} + P_{24})\gamma_8 + (T P_{15} + P_{25})\gamma_9 + (T P_{16} + P_{26})\gamma_{10} \right) +$$

$$\gamma_7 \left(-U_{23} + \frac{P_{33}\gamma_7}{2} + P_{34}\gamma_8 + P_{35}\gamma_9 + P_{36}\gamma_{10} \right) + \gamma_8 \left(-U_{24} + \frac{P_{44}\gamma_8}{2} + P_{45}\gamma_9 + P_{46}\gamma_{10} \right) +$$

$$\gamma_9 \left(-U_{25} + \frac{P_{55}\gamma_9}{2} + P_{56}\gamma_{10} \right) + \gamma_{10} \left(\frac{P_{66}\gamma_{10}}{2} - U_{26} \right) \quad \text{(A1)}$$

with

$$\mathcal{E} = \begin{pmatrix} \gamma_0 & \gamma_1 & \gamma_2 & \gamma_3 & \gamma_4 & \gamma_5 \\ \gamma_2\rho^2 + T\gamma_6 & \gamma_6 & \gamma_7 & \gamma_8 & \gamma_9 & \gamma_{10} \end{pmatrix}. \quad \text{(A2)}$$

The residual for the constraints (15)(a) and (b) active is

$$\frac{W_{11}}{2} + \frac{W_{22}}{2} + \gamma_0 \left(-U_{11} + \frac{P_{11}\gamma_0}{2} + P_{13}\gamma_1 + P_{14}\gamma_2 + P_{15}\gamma_3 + P_{16}\gamma_4 \right) +$$

$$\gamma_2 \left(-U_{14} + \frac{P_{44}\gamma_2}{2} + P_{45}\gamma_3 + P_{46}\gamma_4 \right) + \gamma_3 \left(-U_{15} + \frac{P_{55}\gamma_3}{2} + P_{56}\gamma_4 \right) + \gamma_4 \left(\frac{P_{66}\gamma_4}{2} - U_{16} \right) +$$

$$\gamma_1 \left(-U_{21}\rho^2 + P_{13}\gamma_6\rho^2 + P_{14}\gamma_7\rho^2 + P_{15}\gamma_8\rho^2 + P_{16}\gamma_9\rho^2 - U_{13} + \left(\frac{P_{11}\rho^4}{2} + \frac{P_{33}}{2}\right)\gamma_1 + P_{34}\gamma_2 +\right.$$

$$\left. P_{35}\gamma_3 + P_{36}\gamma_4 + (T P_{11}\rho^2 + P_{12}\rho^2)\gamma_5 \right) + \gamma_5 \left(-T U_{21} - U_{22} + \left(\frac{P_{11}T^2}{2} + P_{12}T + \frac{P_{22}}{2}\right)\gamma_5 +\right.$$

$$\left. (T P_{13} + P_{23})\gamma_6 + (T P_{14} + P_{24})\gamma_7 + (T P_{15} + P_{25})\gamma_8 + (T P_{16} + P_{26})\gamma_9 \right) +$$

$$\gamma_6 \left(-U_{23} + \frac{P_{33}\gamma_6}{2} + P_{34}\gamma_7 + P_{35}\gamma_8 + P_{36}\gamma_9 \right) + \gamma_7 \left(-U_{24} + \frac{P_{44}\gamma_7}{2} + P_{45}\gamma_8 + P_{46}\gamma_9 \right) +$$

$$\gamma_8 \left(-U_{25} + \frac{P_{55}\gamma_8}{2} + P_{56}\gamma_9 \right) + \gamma_9 \left(\frac{P_{66}\gamma_9}{2} - U_{26} \right) \quad \text{(A3)}$$

with

$$\mathcal{E} = \begin{pmatrix} \gamma_0 & 0 & \gamma_1 & \gamma_2 & \gamma_3 & \gamma_4 \\ \gamma_1\rho^2 + T\gamma_5 & \gamma_5 & \gamma_6 & \gamma_7 & \gamma_8 & \gamma_9 \end{pmatrix}. \quad \text{(A4)}$$

The residual for the constraints (15)(a) and (c) active is

$$\frac{W_{11}}{2} + \frac{W_{22}}{2} + \gamma_0 \left(-U_{11} + \frac{P_{11}\,\gamma_0}{2} + P_{12}\,\gamma_1 + P_{13}\,\gamma_2 + P_{14}\,\gamma_3 + P_{15}\,\gamma_4 + P_{16}\,\gamma_5\right) +$$

$$\gamma_1 \left(-U_{12} + \frac{P_{22}\,\gamma_1}{2} + P_{23}\,\gamma_2 + P_{24}\,\gamma_3 + P_{25}\,\gamma_4 + P_{26}\,\gamma_5\right) + \gamma_3 \left(-U_{14} + \frac{P_{44}\,\gamma_3}{2} + P_{45}\,\gamma_4 + P_{46}\,\gamma_5\right) +$$

$$\gamma_4 \left(-U_{15} + \frac{P_{55}\,\gamma_4}{2} + P_{56}\,\gamma_5\right) + \gamma_5 \left(\frac{P_{66}\,\gamma_5}{2} - U_{16}\right) + \gamma_2 \Big(-U_{21}\,\rho^2 + P_{14}\,\gamma_7\,\rho^2 + P_{15}\,\gamma_8\,\rho^2 +$$

$$P_{16}\,\gamma_9\,\rho^2 - U_{13} + \left(\frac{P_{11}\,\rho^4}{2} + \frac{P_{33}}{2}\right)\gamma_2 + P_{34}\,\gamma_3 + P_{35}\,\gamma_4 + P_{36}\,\gamma_5 + (T\,P_{11}\,\rho^2 + P_{12}\,\rho^2)\,\gamma_6\Big) +$$

$$\gamma_6 \left(-T\,U_{21} - U_{22} + \left(\frac{P_{11}\,T^2}{2} + P_{12}\,T + \frac{P_{22}}{2}\right)\gamma_6 + (T\,P_{14} + P_{24})\,\gamma_7 + (T\,P_{15} + P_{25})\,\gamma_8 + (T\,P_{16} + P_{26})\,\gamma_9\right) +$$

$$\gamma_7 \left(-U_{24} + \frac{P_{44}\,\gamma_7}{2} + P_{45}\,\gamma_8 + P_{46}\,\gamma_9\right) + \gamma_8 \left(-U_{25} + \frac{P_{55}\,\gamma_8}{2} + P_{56}\,\gamma_9\right) + \gamma_9 \left(\frac{P_{66}\,\gamma_9}{2} - U_{26}\right)$$

(A5)

with

$$\varepsilon = \begin{pmatrix} \gamma_0 & \gamma_1 & \gamma_2 & \gamma_3 & \gamma_4 & \gamma_5 \\ \gamma_2\,\rho^2 + T\,\gamma_6 & \gamma_6 & 0 & \gamma_7 & \gamma_8 & \gamma_9 \end{pmatrix}.$$

(A6)

The residual for the constraints (15)(a), (b) and (c) active is

$$\frac{W_{11}}{2} + \frac{W_{22}}{2} + \gamma_0 \left(-U_{11} + \frac{P_{11}\,\gamma_0}{2} + P_{13}\,\gamma_1 + P_{14}\,\gamma_2 + P_{15}\,\gamma_3 + P_{16}\,\gamma_4\right) +$$

$$\gamma_2 \left(-U_{14} + \frac{P_{44}\,\gamma_2}{2} + P_{45}\,\gamma_3 + P_{46}\,\gamma_4\right) + \gamma_3 \left(-U_{15} + \frac{P_{55}\,\gamma_3}{2} + P_{56}\,\gamma_4\right) + \gamma_4 \left(\frac{P_{66}\,\gamma_4}{2} - U_{16}\right) +$$

$$\gamma_1 \Big(-U_{21}\,\rho^2 + P_{14}\,\gamma_6\,\rho^2 + P_{15}\,\gamma_7\,\rho^2 + P_{16}\,\gamma_8\,\rho^2 - U_{13} + \left(\frac{P_{11}\,\rho^4}{2} + \frac{P_{33}}{2}\right)\gamma_1 +$$

$$P_{34}\,\gamma_2 + P_{35}\,\gamma_3 + P_{36}\,\gamma_4 + (T\,P_{11}\,\rho^2 + P_{12}\,\rho^2)\,\gamma_5\Big) +$$

$$\gamma_5 \left(-T\,U_{21} - U_{22} + \left(\frac{P_{11}\,T^2}{2} + P_{12}\,T + \frac{P_{22}}{2}\right)\gamma_5 + (T\,P_{14} + P_{24})\,\gamma_6 + (T\,P_{15} + P_{25})\,\gamma_7 + (T\,P_{16} + P_{26})\,\gamma_8\right) +$$

$$\gamma_6 \left(-U_{24} + \frac{P_{44}\,\gamma_6}{2} + P_{45}\,\gamma_7 + P_{46}\,\gamma_8\right) + \gamma_7 \left(-U_{25} + \frac{P_{55}\,\gamma_7}{2} + P_{56}\,\gamma_8\right) + \gamma_8 \left(\frac{P_{66}\,\gamma_8}{2} - U_{26}\right)$$

(A7)

with

$$\varepsilon = \begin{pmatrix} \gamma_0 & 0 & \gamma_1 & \gamma_2 & \gamma_3 & \gamma_4 \\ \gamma_1\,\rho^2 + T\,\gamma_5 & \gamma_5 & 0 & \gamma_6 & \gamma_7 & \gamma_8 \end{pmatrix}.$$

(A8)

FIGURES

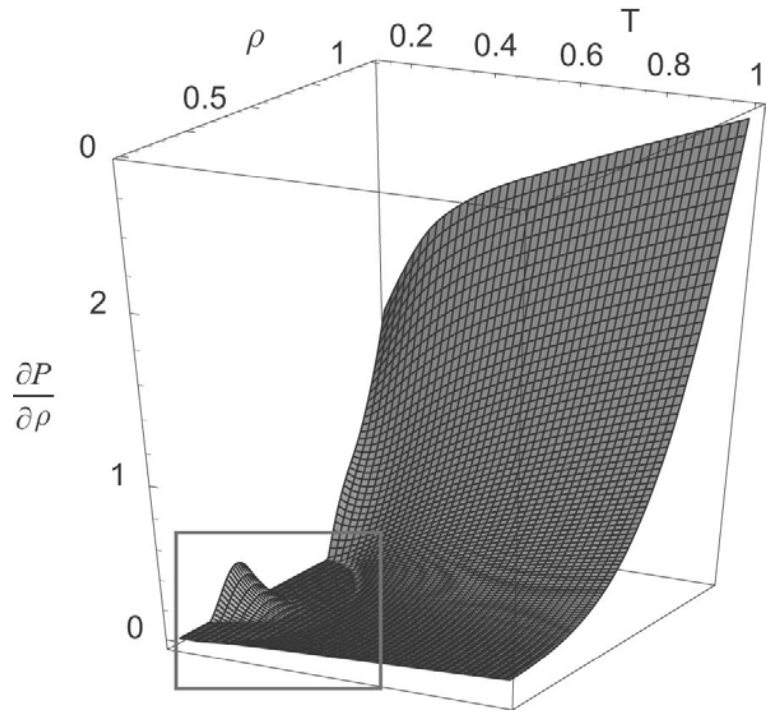

FIG. 1. TRE result for $\partial P/\partial \rho$ with consistency and stability constraints active. Note flat annulus at low temperature caused by activation of stability constraint (box).

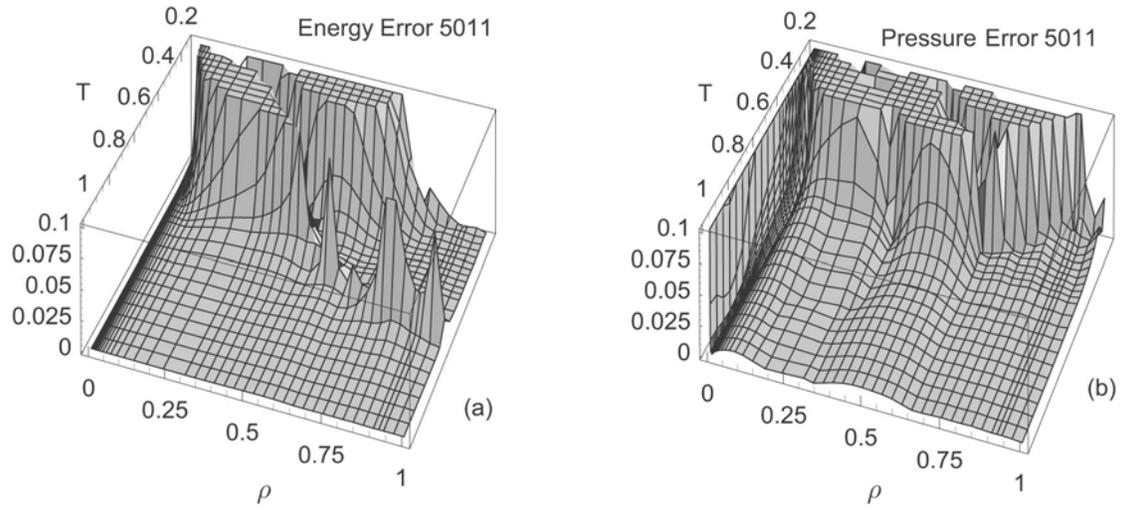

FIG. 2. Error for Oxygen table 5011 when input grid equals output grid. (a) Energy. (b) Pressure.

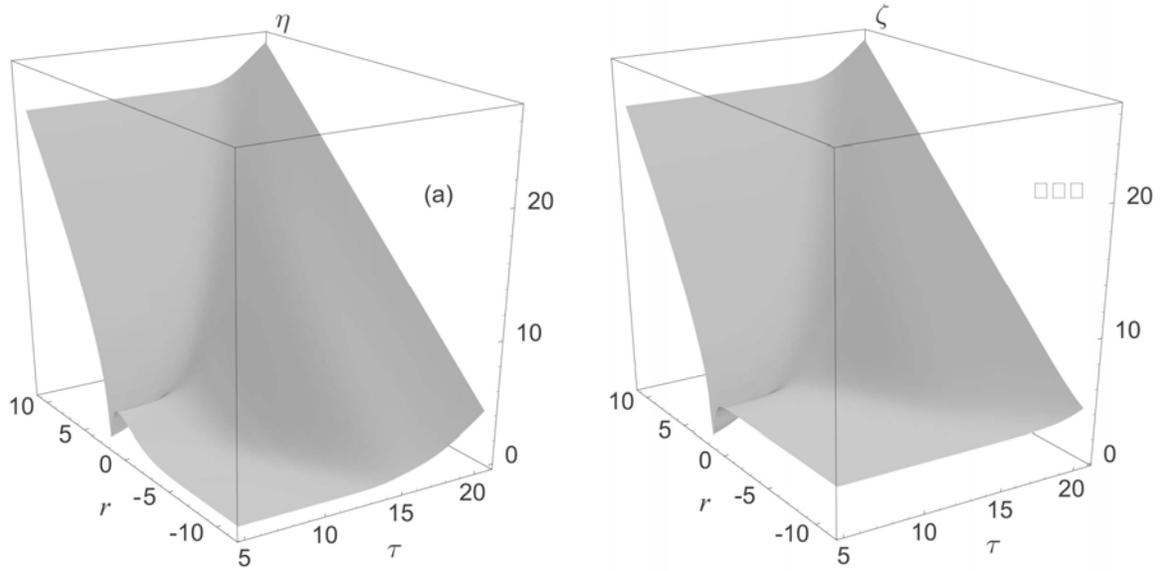

FIG. 3. Results of log-log TRE on table 2984. 75 x 135 grid. (a) Log of shifted energy. (b) Log of shifted pressure.

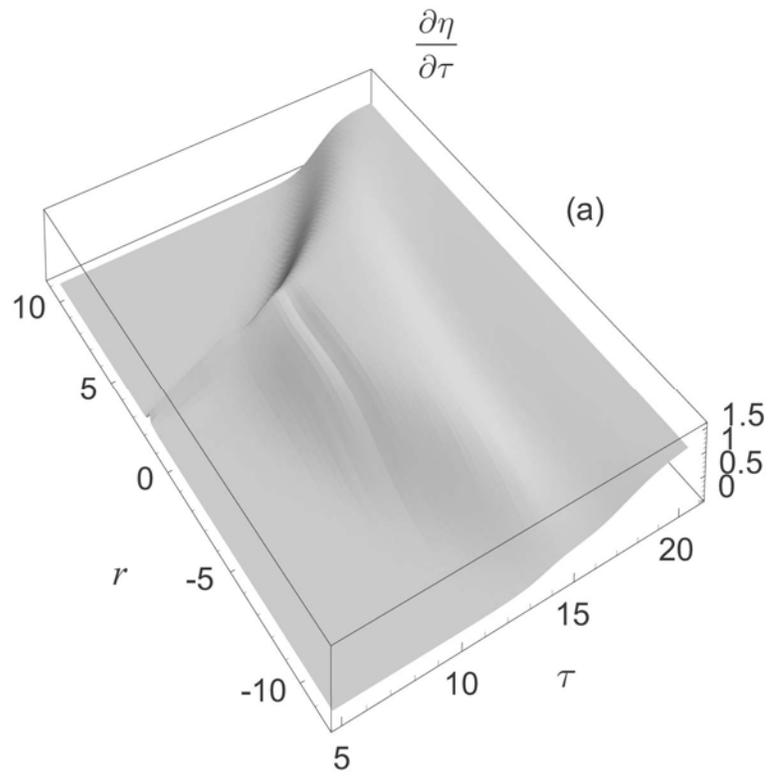

FIG. 4. $\partial\eta/\partial\tau$ from log-log TRE on table 2984.

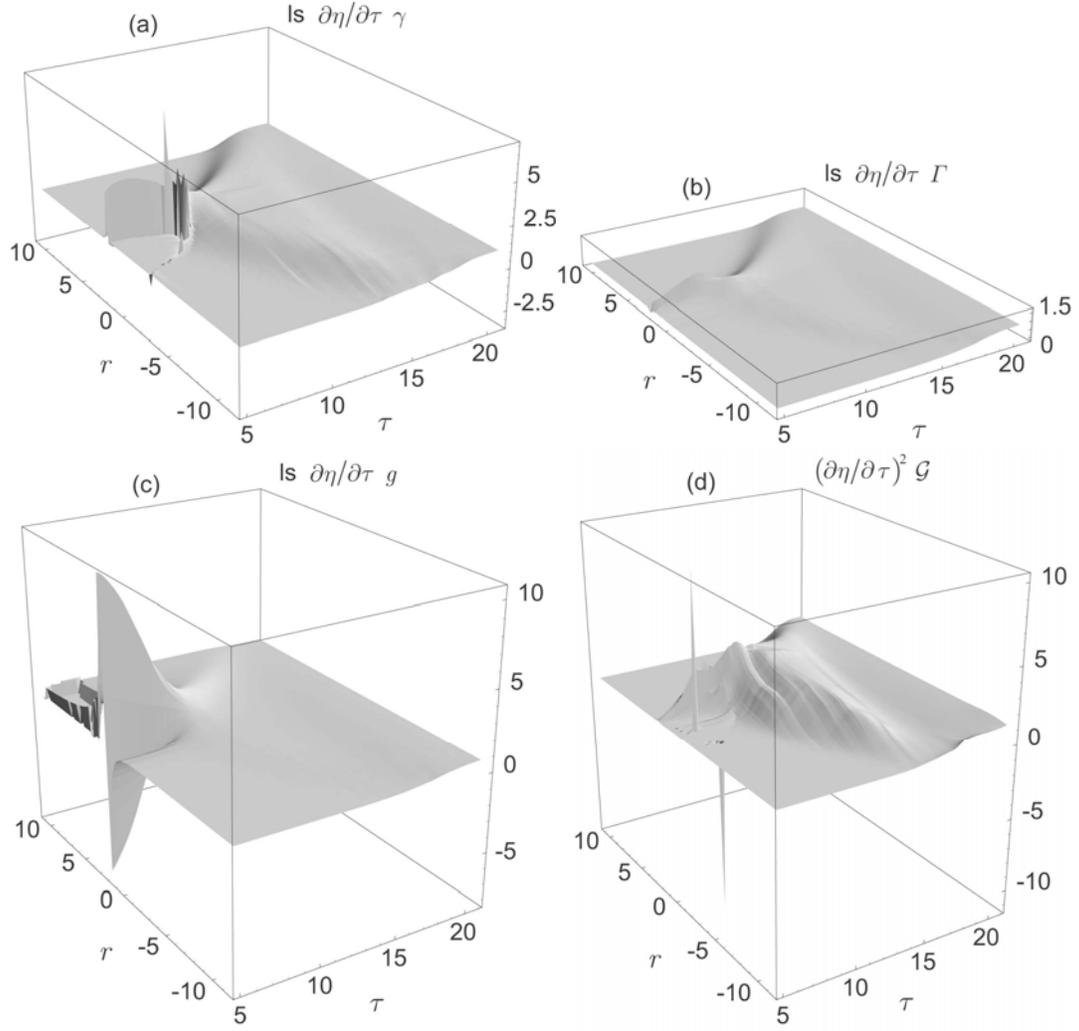

FIG. 5. Dimensionless derivatives by log-log TRE on table 2984 multiplied by $\partial\eta/\partial\tau$ or its square. (a) Log-scale of $\partial\eta/\partial\tau\ \gamma$. (b) Log-scale of $\partial\eta/\partial\tau\ \Gamma$. (c) Log-scale of $\partial\eta/\partial\tau\ g$. (d) Log-scale of $(\partial\eta/\partial\tau)^2\ \mathcal{G}$.

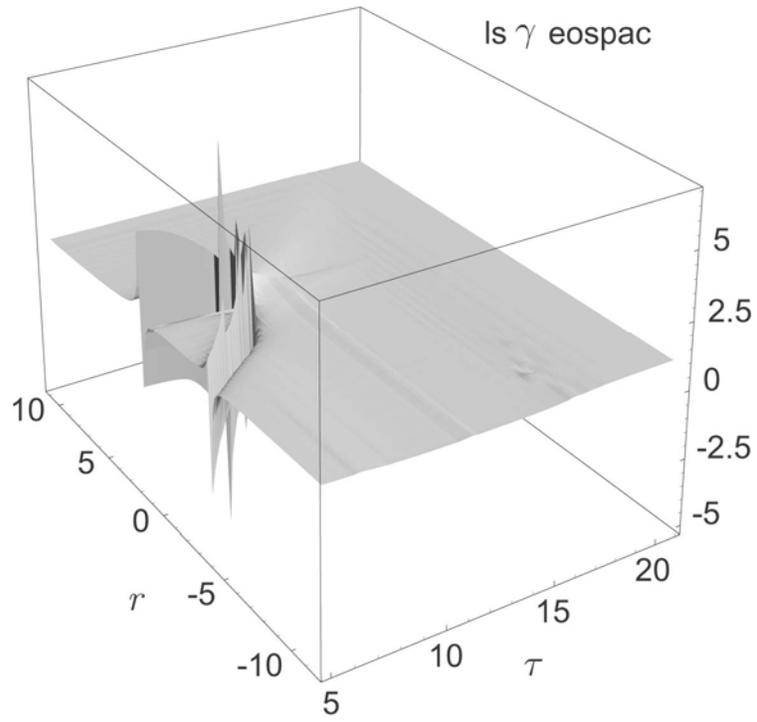

FIG. 6. Plot of $\gamma$ using EOSPAC derivatives for table 2984.

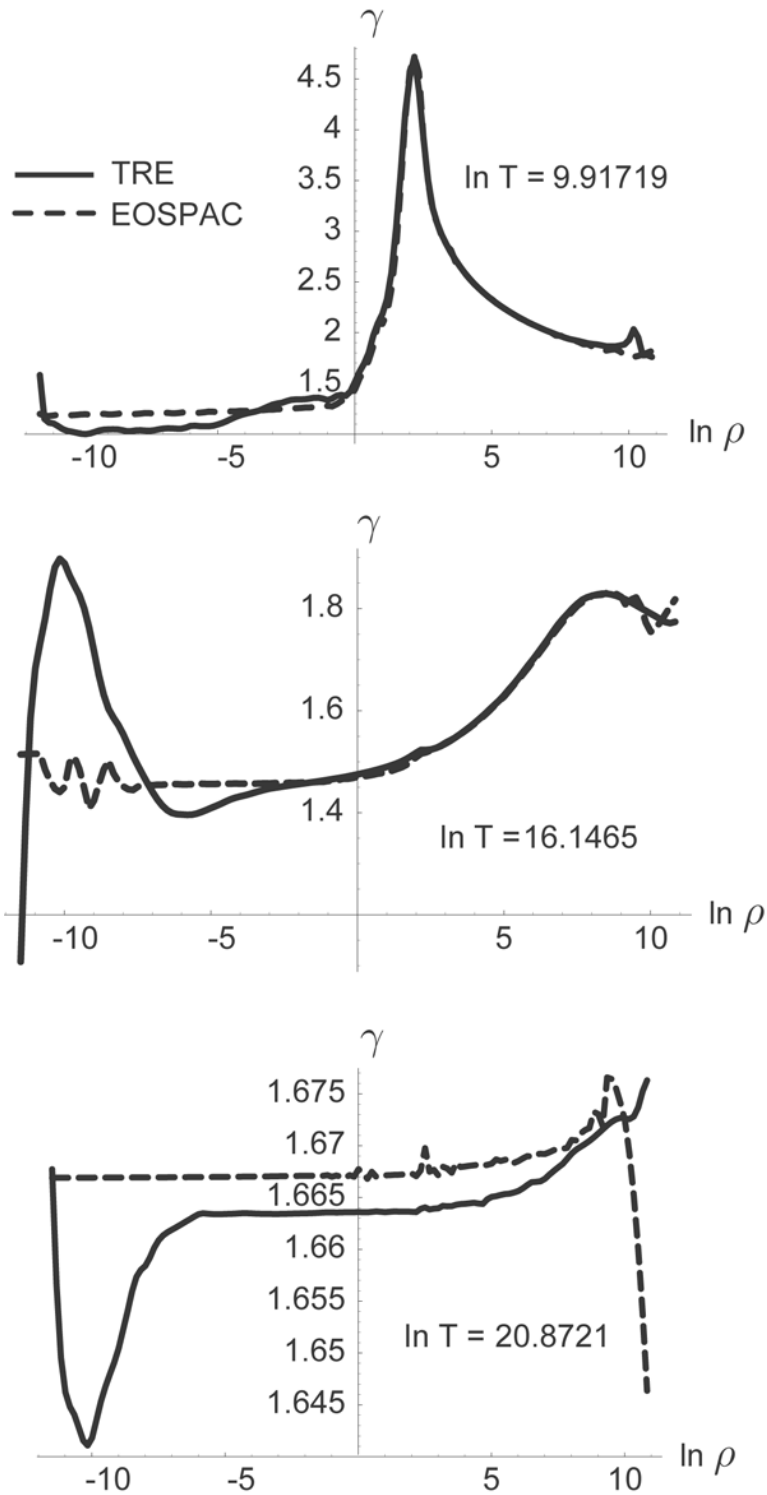

FIG. 7. Comparison of EOSPAC and TRE values of $\gamma$ at $T = 2$ ev, 1 kev, 100 kev.

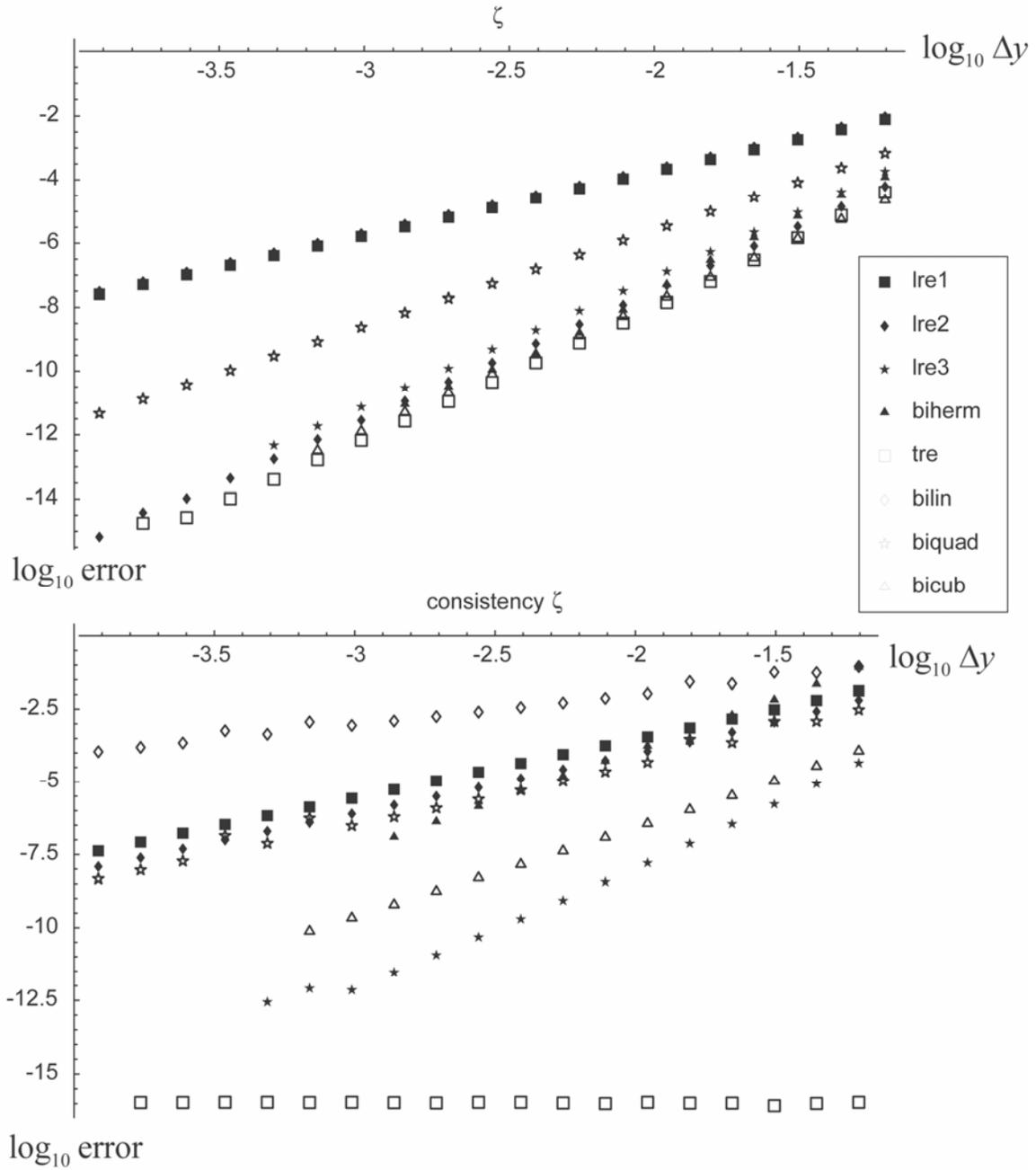

FIG. 8. Error of various methods on analytic EOS of (25) as a function of mesh spacing in the input table for $\zeta$ and consistency.

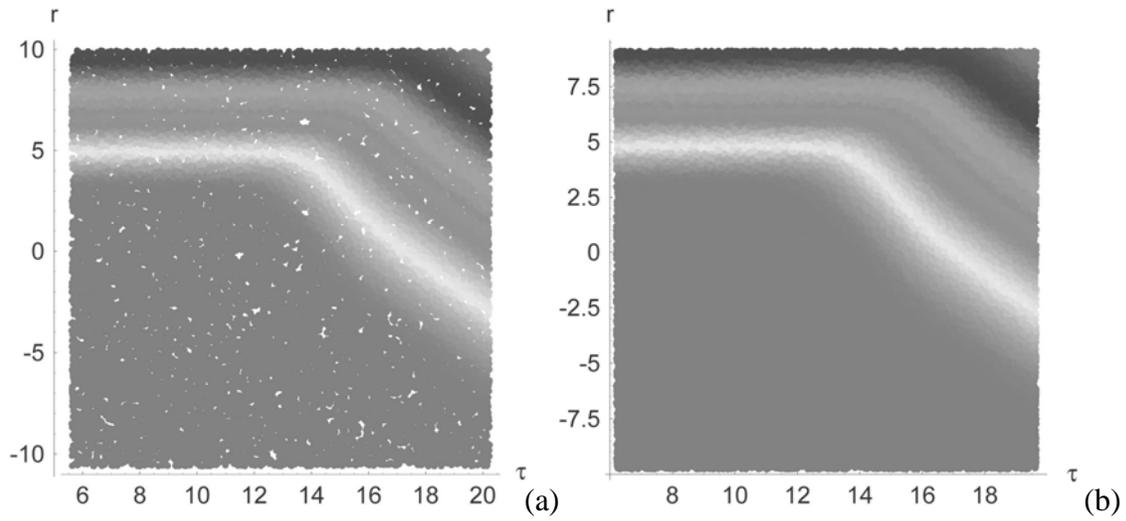

FIG. 9. (a) SESAME table 2984 sampled at 21583 uniformly distributed random points using log-log TRE. (b) Sampling of the data from (a) at 64749 different uniformly distributed random points using log-log TRE. In these examples $\varepsilon_S = P_S = -10^3$. Each point is colored with the value of the interpolant.